# A modified divide-and-conquer based machine learning method for predicting creep life of superalloys

Ronghai Wu[1*], Lei Zeng[1], Xing Ai[2], Yunsong Zhao[3]

[1]School of Mechanics, Civil Engineering and Architecture, Northwestern Polytechnical University, Xi'an, 710129, PR China

[2]Powder Metallurgy Institute, Central South University, Changsha, 410083, PR China

[3]Beijing Institute of Aeronautical Materials, Beijing, 100095, PR China

**Abstract:** Recently Liu et al. (Acta Mater., 2020) proposed a new divide-and-conquer based machine learning method for predicting creep life of superalloys. The idea is enlightening though, the prediction accuracy and intelligence remain to be improved. In the present work, we modify the method by adding a dimensionality reduction algorithm before the clustering step and introducing a grid search algorithm to the regression model selection step. As a consequence, the clustering result becomes much more desirable and the accuracy of predicted creep life is dramatically improved. The root-mean-square error, mean-absolute-percentage error and relevant coefficient of the original method are 0.2341, 0.0595 and 0.9121, while those of the modified method are 0.0285, 0.0196, and 0.9806, respectively. Moreover, the ad-hoc determination of hyperparameters in the original method is replaced by automated determination of hyperparameters in the modified method, which considerably improves the intelligence of the method.

Superalloys are widely used as blades of advanced engines due to their excellent high temperature deformation resistance and oxidation resistance [1-4]. As creep is one of the main deformation modes of superalloys, the investigations of creep behavior and property of superalloys have been flourished using various computational methods [5-8]. Compare with traditional computational methods, machine learning methods do not require presetting of equations for describing material behaviors [9]. Because of that, machine learning is becoming one of the promising methods for the development of superalloys in terms of optimizing process parameters during additive manufacturing [10-12], predicting microstructures [13-21] and mechanical properties [22-25]. Liu et al. recently proposed a new divide-and-conquer based machine learning method which sheds light on predicting mechanical properties of superalloys by artificial intelligence [26]. Although the idea is brilliant, the prediction accuracy and intelligence remain to be improved. To this end, we develop a modified method which will be introduced with the emphasis on modifications compare with the original method, while the details which are same with the original method will not be repeated due to page limitation and can be found in the work of Liu et al. [26].

---

* Corresponding author: ronghai.wu@nwpu.edu.cn (Ronghai Wu) .

The flow chart of the modified divide-and-conquer machine learning method is shown in Fig.1. It starts from the collection of experimental data. We use exactly the same experimental data as the original method, except that a few debatable samples are removed. There are 253 samples in total including 221 samples for training and 32 for testing. Unlike the original method which needs some samples for validation before testing, the modified method does not need validation at all since we use an automated algorithm to determine the hyperparameters. Each sample is described by 27 independent features (i.e. 27 dominations) which include the mass percent of Ni, loading temperature, loading stress, et al. and the corresponding creep rapture life. As different samples may map to different creep mechanisms, to predict the creep life in a more feasible and accurate way, the samples are first divided into different clusters and then treated by different regress models (i.e. the so called "divide-and-conquer"). The original method directly does clustering to the samples. However, we found that direct clustering suffers from the curse of dimensionality. Hence, we propose a dimensionality reduction step before the clustering step to solve the dimensionality curse problem, as shown in the second frame of Fig. 1. Specifically, we use the principal component analysis (PCA) algorithm to reduce dimension from 27 to 2 by:

$$\Sigma = \frac{1}{N} D^{\mathrm{T}} D$$

$$\mathrm{SVD}(\Sigma) = USV^*$$

$$D^{\mathrm{red}} = DU^{\mathrm{red}}$$

where $N$ is the number of samples, $D$ is the experimental data matrix before dimensionality reduction, $\Sigma$ is the covariance matrix, $\text{SVD}(\Sigma)$ is the singular value decomposition of $\Sigma$, $U$ is the unitary matrix whose columns are the eigenvectors of the covariance matrix, $S$ is the nonnegative real diagonal matrix, $V^*$ is the conjugate transpose of $V$, $D^{\text{red}}$ is the experimental data matrix after dimensionality reduction and $U^{\text{red}}$ is a matrix consists of the first two columns of $U$. Note that the dimension after dimensionality reduction dependents on how many first columns that $U^{\text{red}}$ picks up from $U$. Each columns of $U$ has a weight indicating the importance and the weight decrease dramatically from the first to the last column. As the weights of the first two columns (i.e. 0.4 and 0.2) are enough to grasp the main information, we pick up the first two columns. Now the each sample in $D^{\text{red}}$ has two components: principle component one (PC1) and principle component one (PC2). Note that none of them belongs to the 27 features before dimensionality reduction. The original method does not have dimensionality reduction and does clustering based on $D$, while our modified method has dimensionality reduction and does clustering based on $D^{\text{red}}$. Like the original method, we divided the samples by the K-means algorithm, as shown in the third frame of Fig. 1. We additionally introduce a cost function for assessing the clustering results by:

$$J = -\sum_{i=1}^{N}\left(D_i^{\text{red}} - u_{c_i}\right)^2$$

where $c_i$ is the cluster label of sample $D_i^{\text{red}}$ and $u_{c_i}$ is the cluster center that the sample belongs to. $J$ represents the opposite of square distance summation between the samples and cluster centers, bigger $J$ value indicates better clustering results. The data after dimensionality reduction and clustering then enters the step of regression model selection, as shown in the fourth and fifth frames of Fig. 1. We use five regression models: lasso regression (LR), ridge regression (RR), random forest regression (RFR), support vector regression (SVR) and gaussian process regression (GPR), which are the same as the original method. The hyperparameters of regression models in the original method are manually determined by educated guess and adjusted by validation process. Then, the fitness values between all clusters and regression models are calculated and the regression model with maximum fitness value is selected as the optimal regression model (ORM) for the cluster. However, we propose to use the grid search algorithm to automatically determine the hyperparameters, and hence improve the intelligence of the method. The grid search algorithm calculates the relevant coefficient (R-square) at a wide range of hyperparameters for each cluster and regression model, and then offers the hyperparameter with the biggest R-square. The corresponding regression model serves as the optimal model for the cluster. By the processes described above, the whole machine learning method has been set up.

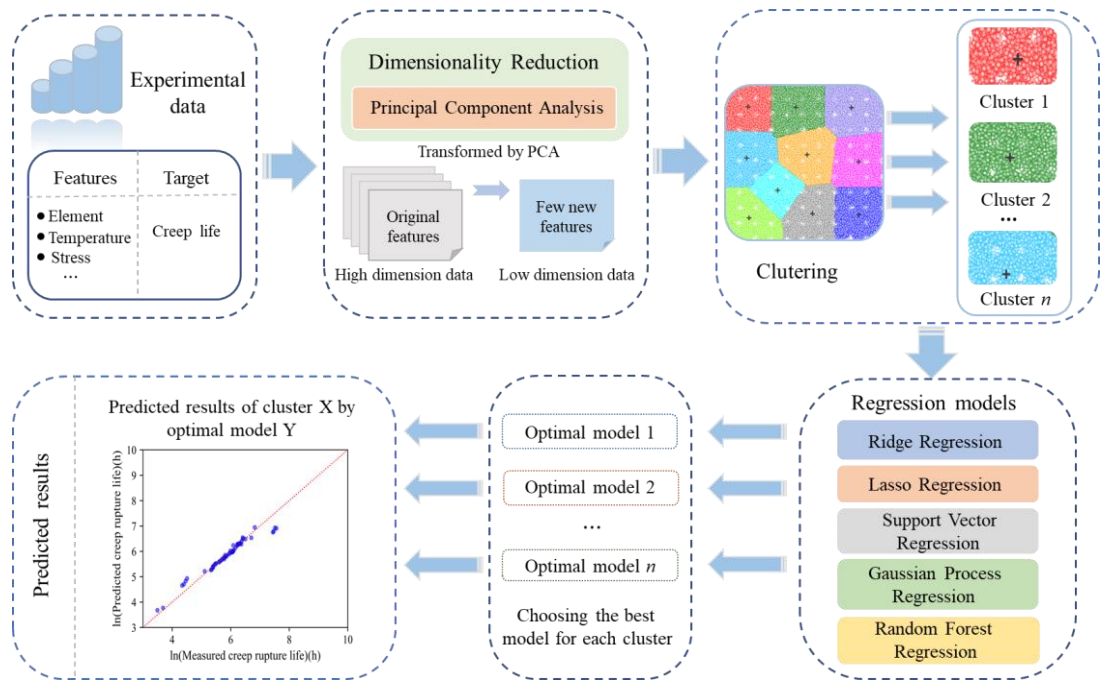

Fig. 1. Flow chart of the modified divide-and-conquer machine learning method.

We implement both the original and modified methods by Python scikit-learn, and compare the results systematically. The clustering results of both methods are shown in Fig. 2, where different colors represent different clusters and the center of $i$th cluster is labeled by number $i$. As can be seen, the clustering result of the modified method is better than that of the original method. More specifically, clusters of the original method considerably overlap with each other (e.g. the cluster 3, 7 2 and 5) while those of the modified method do not. The value of cost function $J$ is -94 for the original method and -7 for the modified method. Note that although the clustering result with respect to PC_1 and PC_2 is plotted in Fig. 1(a) here as well as the Fig.7 in Liu's paper [26], this is for visualization purpose only, the dimensionality reduction does not actually enter the machine learning process in the original method, whereas the dimensionality reduction is an important step in our modified method.

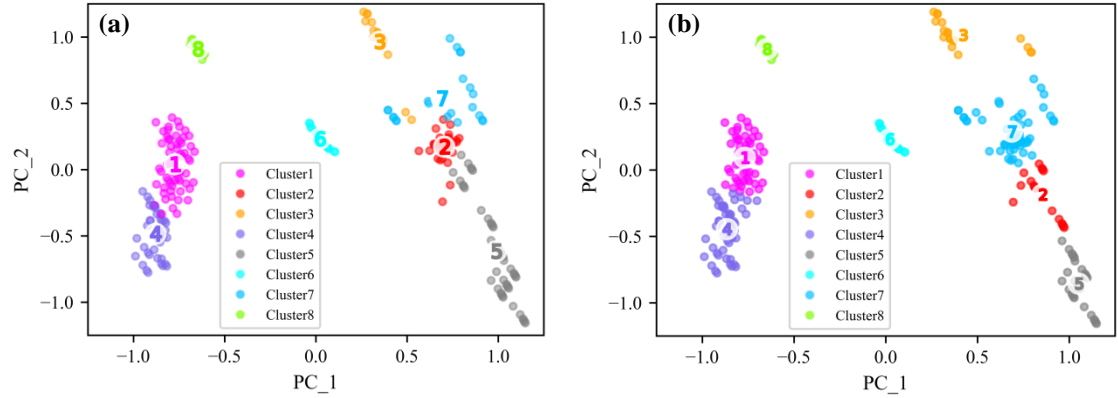

Fig.2. Clustering results with different colors represent different clusters and the center of $i$th cluster is labeled by number $i$: (a) Original method; (b) Modified method.

The selection of optimal models is shown in Table 1. It shows that the ORMs of three clusters (i.e. cluster 4, 5 and 6) are the same for the two methods and those of the rest five clusters are different. For all clusters, the fitness values of the modified method are bigger than those of the original method. The average fitness is 0.9402 for the original method and 0.9804 for the modified method. Besides, the ORM is selected by automatic algorithm rather than educated trial-and-adjust. It is fair to say that the modified method is more accurate and intelligent at determining the optimal model for each cluster.

Table 1 The ORM and corresponding fitness of the original and modified methods

| Cluster | Original method | | Modified method | |
|---|---|---|---|---|
| | ORM | Fitness | ORM | Fitness |
| 1 | RFR | 0.9341 | GPR | 0.9839 |
| 2 | RFR | 0.9673 | SVR | 0.9800 |
| 3 | SVR | 0.9851 | GPR | 0.9934 |
| 4 | RR | 0.9559 | RR | 0.9787 |
| 5 | SVR | 0.9673 | SVR | 0.9775 |
| 6 | SVR | 0.9361 | SVR | 0.9832 |
| 7 | SVR | 0.8542 | RR | 0.9826 |
| 8 | LR | 0.9214 | RR | 0.9641 |
| Average | | 0.9402 | | 0.9804 |

The comparison between experimental and calculated creep rupture life is shown in Fig. 3. As can be seen, the calculated results of the original method are much more scattered than those of the modified method, indicating that the modified method provides better prediction than the original method. To qualitatively compare the experimental and calculated results, the root-mean-square error (RMSE), mean-absolute-percentage error (MAPE) and relevant coefficient (R-square) are defined as:

$$RMSE = \sqrt{\frac{1}{N}\sum_{i=1}^{N}(h(D_i) - y_i)}$$

$$MAPE = \frac{1}{N}\sum_{i=1}^{N}\frac{|h(D_i) - y_i|}{y_i}$$

$$R^2 = \frac{\sum_j(h(D_j) - \frac{1}{N}\sum_{i=1}^{N}y_i)^2}{\sum_j(y_j - \frac{1}{N}\sum_{i=1}^{N}y_i)^2}$$

Where $h(D_i)$ is the experimental creep life of sample $D_i$ and $y_i$ is the predicted creep life. It is the smaller the better for RMSE and MAPE, while the bigger the better for R-square. The RMSE decreases from 0.2341 (original method) to 0.0285 (modified method), MAPE decreases from 0.0595 (original method) to 0.0196 (modified method) and R-square increases from 0.9121 (original method) to 0.9806 (modified method), respectively. The modified divide-and-conquer machine learning method can be a more accurate tool for predicting the creep rapture life of superalloys.

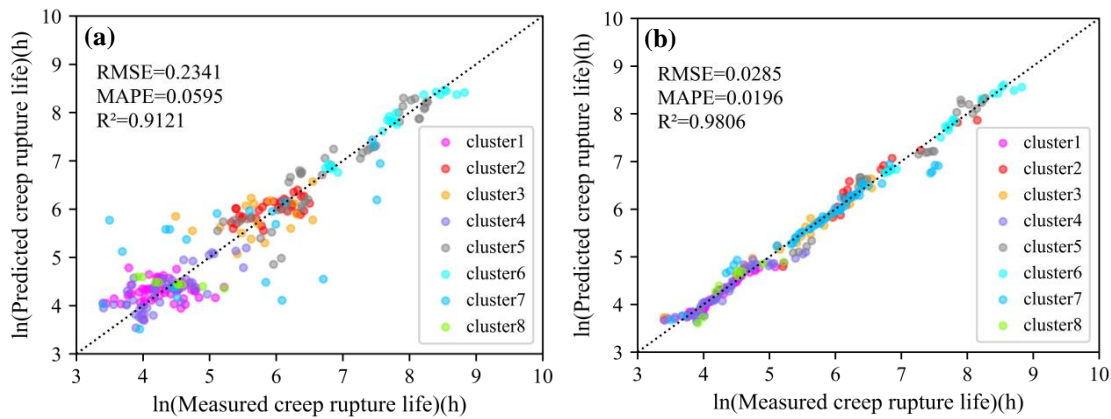

Fig. 3. Experimental and predicted creep rupture life: (a) Original method; (b) Modified method.

We present a modified divide-and-conquer based machine learning method for predicting creep life of superalloys. In the modified mothod, it is demonstrated that the dimensionality reduction is a necessary step before culstering for the divide-and-conquer based machine learning. The cost function for assessing clustering result is -94 for the original method and -7 for the modified method. The RMSE, MAPE and R-square for assessing the prediction accuracy are 0.2341, 0.0595 and 0.9121 for the original method, while 0.0285, 0.0196 and 0.9806 for the modified method, respectively. Besides, the grid search algrithm can provide automatic determination of hyperparameters and coorpesonding optimal model.

**Acknowledgment**

Financial support from National Natural Science Foundation of China (12002275), Natural Science Foundation of Shaanxi Province (2020JQ-125), Hunan Provincial Natural Science Foundation of China (2019JJ50700) and Innovation Foundation of Aero Engine Corporation of China (CXPT-2019-002).